\documentclass[final,3p,times,twocolumn]{elsarticle}
 \biboptions{comma,sort&compress}
\usepackage{here}
 \usepackage{graphicx}
  \usepackage{epsfig}

\def\nin{\noindent}
\def\beq{\begin{equation}}
\def\eeq{\end{equation}}
\def\bea{\begin{eqnarray}}
\def\eea{\end{eqnarray}}

\journal{Nuc. Phys. (Proc. Suppl.)}

\begin{document}

\begin{frontmatter}



\title{Measurements of the proton structure at HERA and their impact for LHC}

 \author[label1,label2]{Alexey Petrukhin\corref{cor1}}
  \address[label1]{DESY, 22607 Notkestr. 85, Hamburg, Germany}
  \address[label2]{SSC RF ITEP, 117218 Bolshaya Cheremushkinskaya 25, Moscow, Russia}
\cortext[cor1]{Speaker on behalf of H1 and ZEUS Collaborations}
\ead{petr@mail.desy.de}



\begin{abstract}
\noindent
Recent results on proton structure functions from H1 and ZEUS Collaborations are presented. 
The data have been recorded in $e^+p$ and $e^-p$ collisions for both
Neutral Current and Charged Current reactions, covering a wide
kinematic range of squared four-momentum transfers $Q^2$, from 0.2\,GeV$^2$ to
30000\,GeV$^2$, and Bjorken $x$ between $\sim$$5*10^{-6}$ and 0.65. Data from both
experiments have been combined, leading to significantly reduced
experimental uncertainties. The combined measurements are analysed in
a NLO QCD fit, and a set of parton density functions, HERAPDF1.0, is
extracted from these data alone.
New direct measurements of the structure function $F_L$, making use of 
dedicated low energy runs of the HERA machine, are also presented.
The impact of the HERA data on the parton density functions and predictions for 
LHC is discussed.

\end{abstract}




\end{frontmatter}


\section{Measurements of the structure function $F_2$}
\nin
A new measurement \cite{1_2000} of deep inelastic lepton-proton scattering (DIS) is based on data collected by the H1 collaboration in the year 2000 with positrons of energy $E_e$=27.6\,GeV and protons of energy $E_p$=920\,GeV, corresponding to a centre-of-mass energy $\sqrt{s}$=319\,GeV. The measurement is performed in the kinematic region of 12\,GeV$^2 \leq Q^2 \leq$ 150\,GeV$^2$ and of $10^{-4} \leq x \leq 0.1$. The luminosity amounts to 22 pb$^{-1}$. This measurement is combined with similar H1 data taken in 1996/97 at $E_p$=820\,GeV \cite{2_1997}. The combined data represent the most precise measurement in the presented kinematic domain with total uncertainties typically in the range of 1.3-2\%.   
The data are used to determine the structure function $F_2(x,Q^2)$, which is observed to rise continuously towards low $x$ at fixed $Q^2$. A NLO QCD analysis is performed to obtain a new set of parton distribution functions H1PDF2009 \cite{1_2000} from the inclusive DIS cross section measurements presented here as well as from previously published H1 measurements at low \cite{3_1999} and high \cite{2_1997} $Q^2$. The data and the NLO QCD fit from H1 data alone are shown in Figure 1.
\begin{figure}[hbt] 
\centerline{\includegraphics[width=8.cm]{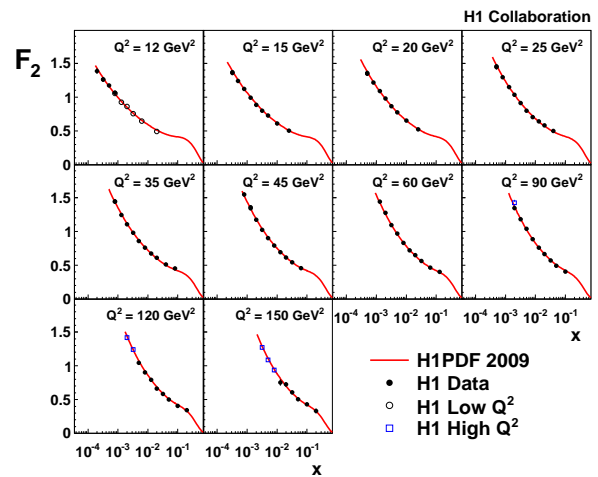}}
\caption{\scriptsize Measurements of the structure function $F_2$ at fixed $Q^2$ as a function of $x$. The new data (closed circles) are complemented by the previously published data at low $Q^2$ (open circles) \cite{3_1999} and high $Q^2$ (open boxes) \cite{2_1997}. The error bars represent the total measurement uncertainties. The solid curve represents the NLO QCD fit to H1 data alone.}
\label{fig1} 
\end{figure} 
\nin

\nin
\section{Combined H1 and ZEUS measurements}
\nin
A combination \cite{4_comb} is presented of the inclusive DIS cross sections measured by the
H1 and ZEUS Collaborations in neutral current unpolarised $ep$ scattering at
HERA during the period 1994-2000. The luminosity amounts to 120\,pb$^{-1}$ per experiment. The data cover several orders of magnitude in $Q^2$, and in Bjorken $x$. The combination method used
takes the correlations of systematic uncertainties into account, resulting in an improved
accuracy. 
The input data from H1 and ZEUS are consistent with each other
at $\chi^2/$ndf = 636.5/656. The total uncertainty of the combined data set is 1\%
in the best measured region, 20\,GeV$^2 < Q^2 < $ 100\,GeV$^2$. Figure 2 shows the combined HERA results: scaling violations predicted by the theory of QCD and the HERAPDF1.0 fit which will be discussed in the next section.     
\begin{figure}[hbt] 
\centerline{\includegraphics[width=8.cm]{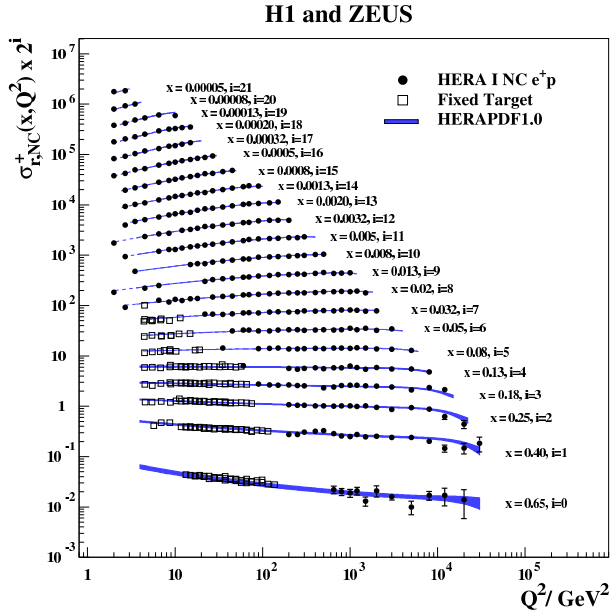}}
\caption{\scriptsize HERA combined neutral current reduced cross section \cite{4_comb} and fixed-target data compared to the HERAPDF1.0 fit. The bands represent the total uncertainty of the fit.}
\label{fig2} 
\end{figure} 
\begin{figure}[hbt] 
\centerline{\includegraphics[width=9.cm]{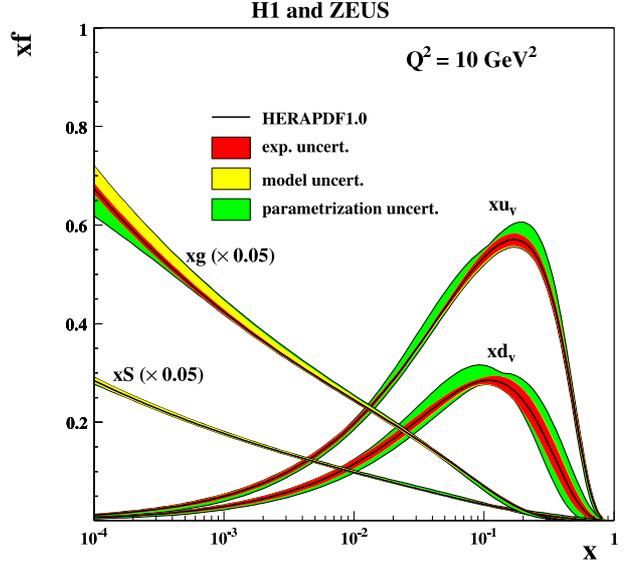}}
\caption{\scriptsize The parton distribution functions from the HERAPDF1.0 at $Q^2$ = 10\,GeV$^2$. The gluon and sea distributions are scaled
down by a factor 20. The experimental, model and parametrisation uncertainties are shown separately (see \cite{4_comb}).}
\label{fig3} 
\end{figure} 
\nin
\section{QCD analysis of the combined data}
\nin
The combined data sets on inclusive neutral and charged current $e^+p$ and $e^-p$ cross sections measured at HERA during the period 1994-2000 are the sole
input to a next-to-leading order QCD analysis which determines
a set of parton distributions, HERAPDF1.0, with small experimental uncertainties. This
set includes an estimate of the model and parametrisation uncertainties of the fit result as explained in \cite{4_comb}. The HERAPDF1.0 fit results are shown in Figures 2 and 3. Scaling violations in Figure 2 are well described over four orders of magnitude in $x$ and $Q^2$ by the HERAPDF1.0 fit with a $\chi^2/$ndf = 532/582. In spite of the fact that fixed target data were not included into the fit, they are described by HERAPDF1.0 as well. Due to the high precision of the combined data set, the parametrisation HERAPDF1.0 has total uncertainties at the level of a few percent at low $x$.
\nin
\section{Measurements of the structure function\,$F_L$}
\nin
Figures 4 and 5 show the first direct measurements of the structure function\,$F_L$ performed by the H1 \cite{5_flh1} and ZEUS \cite{6_flzeus} Collaborations. The H1 Collaboration has also performed preliminary measurements of the structure function\,$F_L$ at low $2.5 \leq Q^2 \leq 25$\,GeV$^2$ by using of Backward Silicon Tracker data [7] and high $35 \leq Q^2 \leq 800$\,GeV$^2$ by using of Liquid Argon calorimeter data [8]. The measurement of $F_L$ requires several sets of DIS cross sections at fixed $x$ and $Q^2$ but at different values of inelasticity $y$. This was achieved at HERA by variation of the proton beam energy whilst keeping the lepton beam energy fixed. The current measurements are
based on inclusive deep inelastic $e^+p$ scattering cross section measurements with a positron beam energy of 27.6\,GeV and proton beam energies of 920, 575 and 460\,GeV. Employing the energy dependence of the cross section, $F_L$ is measured in the range of $12 \leq Q^2
 \leq 130$\,GeV$^2$ and low Bjorken $x$ of $0.00024 \leq x \leq 0.007$. The $F_L$ values agree in this kinematic domain with higher order QCD calculations based on
 parton densities obtained using cross section data previously measured at HERA.
\begin{figure}[hbt] 
\centerline{\includegraphics[width=8.cm]{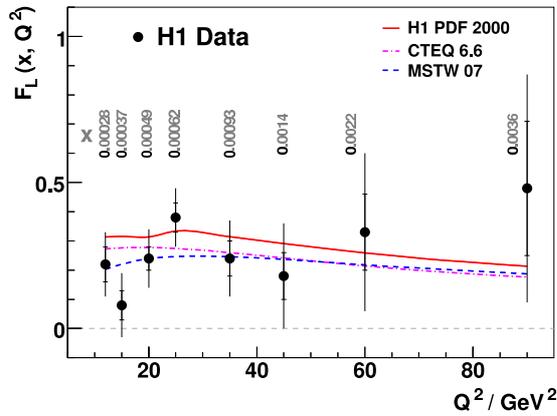}}
\caption{\scriptsize The structure function $F_L$ from the H1 data \cite{5_flh1} compared to the different theoretical predictions. The full error bars
include the statistical and systematic uncertainties added in quadrature.}
\label{fig4} 
\end{figure} 
\begin{figure}[hbt] 
\centerline{\includegraphics[width=8.cm]{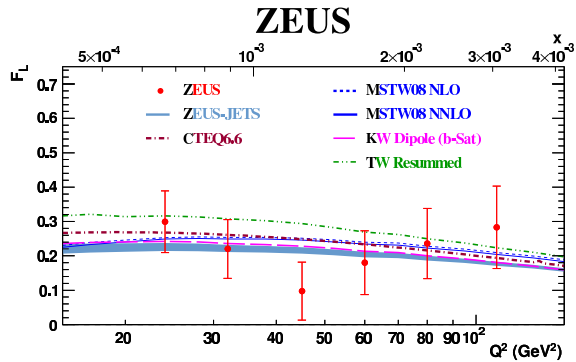}}
\caption{\scriptsize The structure function $F_L$ from the ZEUS data \cite{6_flzeus} compared to the different theoretical predictions. The full error bars
include the statistical and systematic uncertainties added in quadrature.}
\label{fig5} 
\end{figure} 
\nin
Figure 6 shows a new preliminary combined HERA measurement of the structure function $F_L$ [9]. The measurements of $F_L$ use different parts of the H1 and ZEUS detectors covering when combined a wide range of squared four-momentum transfers $2.5 \leq Q^2 \leq 800~$GeV$^2$ and Bjorken $x$ between 0.00005 and 0.035. The data are compared with higher order QCD prediction based on HERAPDF1.0 and found to be in a good agreement with this fit except the lowest $Q^2$ region. 
\begin{figure}[hbt] 
\centerline{\includegraphics[width=9.cm]{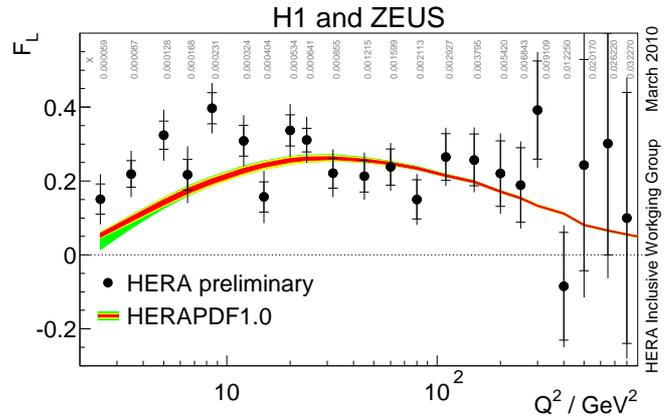}}
\caption{\scriptsize The structure function $F_L$ from the H1 and ZEUS data\,[9] in extended kinematic range of squared four-momentum transfers $2.5 \leq Q^2 \leq 800~$GeV$^2$ and $0.00005 \leq x \leq 0.035$. The data are compared to NLO QCD fit HERAPDF1.0. The full error bars include the statistical and systematic uncertainties added in quadrature.}
\label{fig6} 
\end{figure} 
\nin
\section{HERAPDF predictions for physics at the LHC}
\nin
One
important ingredient for precise cross section predictions at the LHC are the parton
distribution functions of the proton. They can be obtained from global fits to DIS data. In this section we present the HERAPDF1.0 predictions for the important cross sections at the LHC. At the moment the LHC experiments are collecting data at a center-of-mass energy of 7\,TeV and they need an accurate predictions for cross sections and uncertainties. Figures 7,8 show the cross
section for a light Higgs boson with mass $m_H$=120\,GeV. Light Higgs production
originates predominatly for gluons, which are measured at HERA from scaling violation. The prediction has an experimental uncertainty at the level of 2-3\% in a wide range of rapidity $y$. Figures 9,10 show the prediction and the uncertainties for the sea quark dominated process. The prediction has an experimental uncertainty smaller than 1\% and total uncertainty at the level of 5\% or smaller for rapidity $y<2.5$. With the precision quoted here this process can be used as a luminosity monitor for the LHC machine.          
\nin
\begin{figure}[hbt] 
\centerline{\includegraphics[width=8.cm]{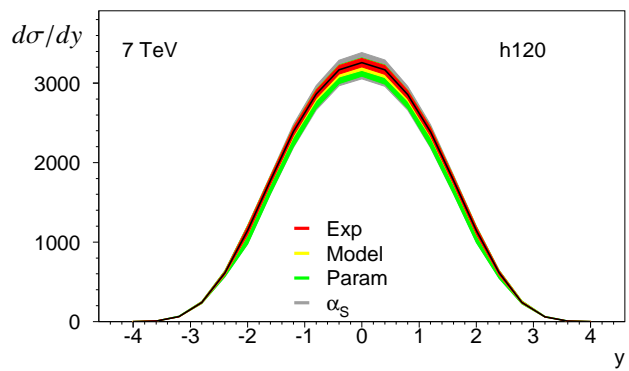}
\put(-235,130){$d\sigma / dy$}}
\caption{\scriptsize The HERAPDF1.0 prediction for the differential cross section $d\sigma / dy$ of the process $gg\rightarrow H$ for a Higgs mass of 120\,GeV as a function of rapidity $y$ for the LHC data at 7\,TeV. Different sources of uncertainty are shown separately.}
\label{fig9} 
\end{figure} 
\nin
\begin{figure}[hbt] 
\centerline{\includegraphics[width=8.cm]{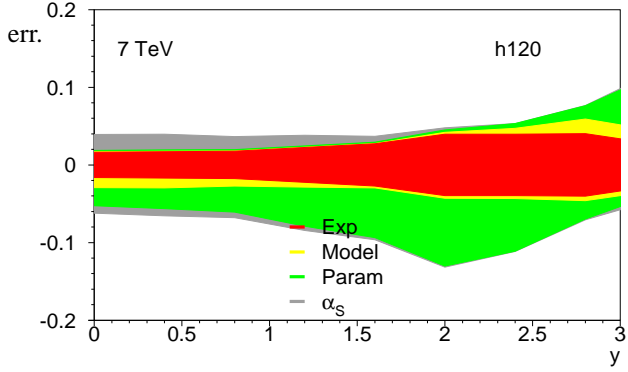}
\put(-235,130){err.}}
\caption{\scriptsize The HERAPDF1.0 prediction uncertainties for the differential cross section $d\sigma / dy$ of the process $gg\rightarrow Higgs$ for a Higgs mass of 120\,GeV as a function of rapidity $y$ for the LHC data at 7\,TeV.}
\label{fig10} 
\end{figure} 
\nin
\begin{figure}[hbt] 
\centerline{\includegraphics[width=8.cm]{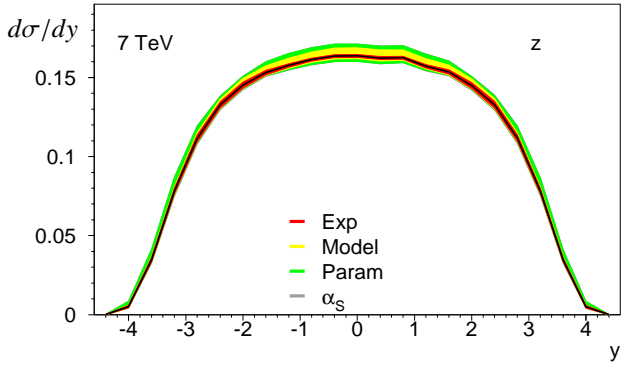}
\put(-235,130){$d\sigma / dy$}}
\caption{\scriptsize The HERAPDF1.0 prediction for the differential cross section $d\sigma / dy$ of the process $Z\rightarrow ee$ as a function of rapidity $y$ for the LHC data at 7\,TeV. Different sources of uncertainty are shown separately.}
\label{fig7} 
\end{figure} 
\nin
\begin{figure}[hbt] 
\centerline{\includegraphics[width=8.cm]{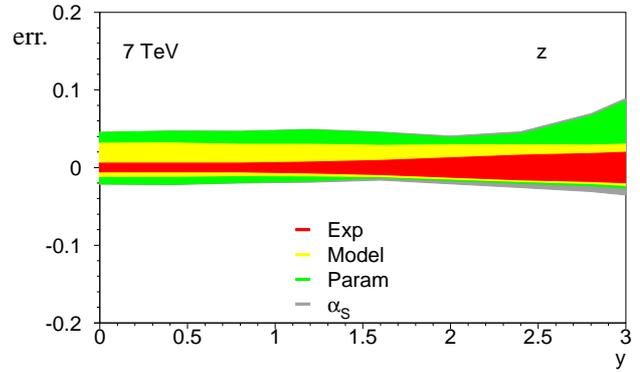}
\put(-235,130){err.}}
\caption{\scriptsize The HERAPDF1.0 prediction uncertainties for the differential cross section $d\sigma / dy$ of the process $Z\rightarrow ee$ as a function of rapidity $y$ for the LHC data at 7\,TeV.}
\label{fig8} 
\end{figure} 
\nin
\section*{Acknowledgements}
\nin
I would like to thank the organizers of QCD 2010 conference and the LPTA-Montpellier for hospitality.












\end{document}